\documentclass[preprint,showkeys,showpacs,preprintnumbers,amsmath,amssymb,tightenlines]{revtex4}
\usepackage{amsfonts}
\usepackage{graphicx}
\usepackage{natbib}
\usepackage{dcolumn}% Align table columns on decimal point
\usepackage{bm}% bold math

\begin{document}

\preprint{APFA06}

\title{How Do Output Growth Rate Distributions Look Like?\\Some Time-Series Evidence on OECD Countries\\}% Force line breaks with \\

\author{Giorgio Fagiolo}
 \altaffiliation{Corresponding Author. University of Verona, Italy and Sant'Anna School of Advanced Studies, Pisa, Italy.
 Mail address: Sant'Anna School of Advanced Studies, Piazza Martiri della Libert\`{a} 33, I-56127 Pisa, Italy.
 Tel: +39-050-883343 Fax: +39-050-883344}
 \email{giorgio.fagiolo@univr.it}

\author{Mauro Napoletano}
\altaffiliation{Chair of Systems Design, ETH Zurich, 8032 Zurich,
  Switzerland and Sant'Anna School
of Advanced Studies, Pisa, Italy.} \email{mnapoletano@ethz.ch}

\author{Andrea Roventini}
 \altaffiliation{University of Modena and Reggio Emilia, Italy and Sant'Anna School of Advanced Studies, Pisa, Italy.}
\email{aroventini@sssup.it}

\date{July 2006}

\begin{abstract}
\noindent This paper investigates the statistical properties of
within-country GDP and industrial production (IP) growth rate
distributions. Many empirical contributions have recently pointed
out that cross-section growth rates of firms, industries and
countries all follow Laplace distributions. In this work, we test
whether also within-country, time-series GDP and IP growth rates
can be approximated by tent-shaped distributions. We fit output
growth rates with the exponential-power (Subbotin) family of
densities, which includes as particular cases both the Gaussian
and the Laplace distributions. We find that, for a large number of
OECD countries including the U.S., both GDP and IP growth rates
are Laplace distributed. Moreover, we show that fat-tailed
distributions robustly emerge even after controlling for outliers,
autocorrelation and heteroscedasticity.
\end{abstract}

\keywords{Output Growth Rate Distributions, Laplace Distribution,
Cross-Country Analysis, Time Series, Output Dynamics.}

\pacs{89.65.Gh; 89.90.+n; 02.60.Ed}

\maketitle

\section{Introduction\label{Section:Introduction}}

In recent years, empirical cross-section growth rate distributions
of diverse economic entities (i.e., firms, industries and
countries) have been extensively explored by both economists and
physicists \cite{Sea96,
Cea98,Lea98,Aea97,BnS03a,BnS03b,CnD04,Fuetall05,Bea05,SnT06}.

The main result of this stream of literature was that, no matter
the level of aggregation, growth rates tend to
\textit{cross-sectionally} distribute according to densities that
display tails fatter than those of a Gaussian distribution. From
an economic point of view, this implies that growth patterns tend
to be quite lumpy: large growth events, no matter if positive or
negative, seem to be more frequent than what a Gaussian model
would predict.

For example, at the microeconomic level, growth rates of U.S.
manufacturing firms (pooled across years) appear to distribute
according to a Laplace\cite{Sea96,Aea97}. This result robustly
holds even if one disaggregates across industrial sectors and/or
considers cross-section distributions in each given year
\cite{BnS03a,BnS03b}. Moreover, in some countries (e.g., France)
firm growth rates display tails even fatter than those of a
Laplace density \cite{Bea05}. Interestingly, similar findings are
replicated at higher aggregation levels: both growth rates of
industrial sectors \cite{CnD04,SnT06} and countries
\cite{Cea98,Lea98,CnD04} display tent-shaped patterns.

Existing studies have been focusing only on \textit{cross-section}
distributions. In this paper, on the contrary, we ask whether
fat-tailed distributions also emerge \textit{across time within a
single country}. More precisely, for any given country, we
consider GDP and industrial production (IP) time series and we
test whether their growth rate distributions can be well
approximated by densities with tails fatter than the Gaussian
ones.

Our analysis shows that in the U.S. both GDP and IP growth rates
distribute according to a Laplace. Similar results hold for a
large sample of OECD countries. Interestingly enough, this
evidence resists to the removal of outliers, heteroscedasticity
and autocorrelation from the original time series. Therefore,
fat-tails emerges as a inherent property of output growth
residuals, i.e. a fresh stylized fact of output dynamics.

Our work differs from previous, similar ones
\cite{Cea98,Lea98,CnD04} in a few other respects. \textit{First},
we depart from the common practice of using annual data to build
output growth rate distributions. We instead employ monthly and
quarterly data. This allows us to get longer series and better
appreciate their business cycle features. \textit{Second}, we fit
output growth rates with the exponential-power (Subbotin)
distribution \cite{Subbotin}, which encompasses Laplace and
Gaussian distributions as special cases. This choice allows us to
measure how far empirical growth rate distributions are from the
Normal benchmark \footnote{A thorough comparative study on the
goodness-of-fit of the Subbotin distribution vis-\`{a}-vis
alternative fat-tails distributions (e.g., Student's-t, Cauchy,
Levy-Stable) is the next point in our agenda.}. \textit{Finally},
we check the robustness of our results to the presence of
outliers, heteroscedasticity and autocorrelation in output growth
rate dynamics.

The paper is organized as follows. In Section
\ref{Section:DataMeth} we describe our data and the methodology we
employ in our analysis. Empirical results are presented in Section
\ref{Section:Results}. Finally, Section \ref{Section:Conclusions}
concludes.

\section{Data and Methodology\label{Section:DataMeth}}

Our study employs two sources of (seasonally adjusted) data. As
far as the U.S. are concerned, we employ data drawn from the FRED
database. More specifically, we consider quarterly real GDP
ranging from $1947Q1$ to $2005Q3$ (235 observations) and monthly
IP from $1921M1$ to $2005M10$ (1018 observations). Analyses for
the OECD sample of countries are performed by relying on monthly
IP data from the ``OECD Historical Indicators for Industry and
Services'' database ($1975M1 - 1998M12$, 287 observations).

The main object of our analysis is output growth rate $g(t)$,
defined as:
\begin{equation}
g(t)=\frac{Y(t)-Y(t-1)}{Y(t-1)}\cong y(t)-y(t-1)=dy(t), \label{eq:growth_rates}
\end{equation}
where $Y(t)$ is the output level (GDP or IP) at time $t$ in a
given country, $y(t)=log[Y(t)]$ and $d$ is the first-difference
operator.

Let $T_n=\{t_1,...,t_n\}$ the time interval over which we observe
growth rates. The distribution of growth rates is therefore
defined as $G_T=\{g(t),t\in T_n\}$. We study the shape of
$G_{T_n}$ in each given country following a parametric approach.
More precisely, we fit growth rates with the exponential-power
(Subbotin) family of densities \footnote{More on fitting Subbotin
distributions to economic data is in \cite{BnS03a,BnS03b}.}, whose
functional form reads:
\begin{equation}
f(x)=\frac{1}{2ab^{\frac{1}{b}}\Gamma(1+\frac{1}{b})}
e^{-\frac{1}{b}|\frac{x-m}{a}|^{b}}, \label{eq:subbotin}
\end{equation}
where $a>0$, $b>0$ and $\Gamma(\cdot)$ is the Gamma function. The
Subbotin distribution is thus characterized by three parameters: a
\textit{location} parameter $m$, a \textit{scale} parameter $a$
and a \textit{shape} parameter $b$. The location parameter
controls for the mean of the distribution. Therefore it is equal
to zero up to a normalization that removes the average growth
rate. The scale parameter is proportional to the standard
deviation.

% Fig 1
\begin{figure}[tbp]
\centering {\includegraphics[width=8cm]{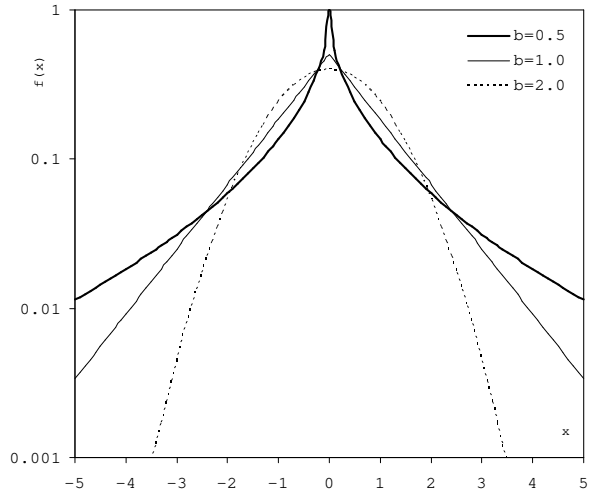}}
\caption{The exponential-power (Subbotin) density for $m=0$, $a=1$
and different shape parameter values: (i) $b=2$: Gaussian density;
(ii) $b=1$: Laplace density; (iii) $b=0.5$: Subbotin with
super-Laplace tails. Note: Log scale on the y-axis.}
\label{Fig:subbodens}
\end{figure}

The shape parameter is the crucial one for our aims, as it
directly gives information about the fatness of the tails: the
larger $b$, the thinner are the tails. Note that if $b=1$ the
distribution reduces to a Laplace, whereas for $b=2$ we recover a
Gaussian. Values of $b$ smaller than one indicate super-Laplace
tails (see Figure \ref{Fig:subbodens} for an illustration).

In our exercises, we fit empirical distributions $G_{T_N}$ with
the Subbotin density (\ref{eq:subbotin}) by jointly estimating the
three parameters by a standard maximum likelihood procedure (see
\cite{Subbo04} for details).

\section{Empirical Results\label{Section:Results}}

In this section we present our main empirical results. We begin with an
analysis of U.S. growth rate distributions. Next, we extend our results to
other OECD countries. Finally, we turn to a robustness analysis of growth
residuals, where we take into account the effects of outliers,
heteroscedasticity and autocorrelation.

\subsection{Exploring U.S. Output Growth Rate Distributions\label{Subsection:US}}

Let us start by some descriptive statistics on U.S. output growth rates. Table
\ref{Tab:ss_us} reports the first four moments of U.S. time series. Standard
deviations reveal that after World War II growth rates of industrial production
and GDP have been characterized by similar volatility levels.
\begin{table*}
\caption{U.S. Output Time Series: Summary Statistics. Asterisk
(*): Significant at 5\% level.} \label{Tab:ss_us}
\begin{ruledtabular}
\begin{tabular}{lcccccc}
{\bf Series} & {\bf Mean} & {\bf Std. Dev.} & {\bf Skewness} & {\bf Kurtosis} & {\bf Jarque-Bera } & {\bf Lilliefors } \\
&  &  &  &   & {\bf  test} & {\bf  test} \\
\hline
       GDP &     0.0084 &     0.0099 &    -0.0891 &     4.2816 &   15.4204* &    0.0623* \\
 IP (1921) &     0.0031 &     0.0193 &     0.3495 &    14.3074 & 5411.7023* &    0.1284* \\
 IP (1947) &     0.0028 &     0.0098 &     0.3295 &     8.1588 &  784.0958* &    0.0822* \\
\end{tabular}
\end{ruledtabular}
\end{table*}
The standard deviation of IP growth rates becomes higher if the
series is extended back to 1921. Skewness is close to zero: -0.09
for GDP and $\approx0.33$ for IP. Notice that both the Jarque-Bera
and Lilliefors normality tests reject the hypothesis that our
series are normally distributed. Furthermore, the relatively high
reported kurtosis values suggest that output growth rate
distributions display tails fatter than the Gaussian distribution.
In order to better explore this evidence, we fit U.S. output
growth rates distribution with the Subbotin density (see eq.
\ref{eq:subbotin}).

\begin{table}[h]
\caption{U.S. Output Growth Rate Distribution: Estimated Subbotin
Parameters.} \label{Tab:US}
\begin{ruledtabular}
\begin{tabular}{lcccccc}
{\bf }  & \multicolumn{ 6}{c}{{\bf Estimated Parameters}} \\
%\hline
    {\bf } & \multicolumn{ 2}{c}{{$\widehat{b}$}} & \multicolumn{ 2}{c}{{\bf $\widehat{a}$}} & \multicolumn{ 2}{c}{{\bf $\widehat{m}$}} \\
    {\bf Series} & {Par.} & {Std. Err.} & {Par.} & {Std. Err.} & {Par.} & {Std. Err.} \\
\hline
       GDP &     1.1771 &     0.1484 &     0.0078 &     0.0006 &     0.0082 &     0.0006 \\
 IP (1921) &     0.6215 &     0.0331 &     0.0091 &     0.0004 &     0.0031 &     0.0002 \\
 IP (1947) &     0.9940 &     0.0700 &     0.0068 &     0.0003 &     0.0030 &     0.0003 \\
 \end{tabular}
\end{ruledtabular}
\end{table}

\begin{figure}[h]
\begin{minipage}[h]{7.5cm}
\centering {\includegraphics[width=6cm]{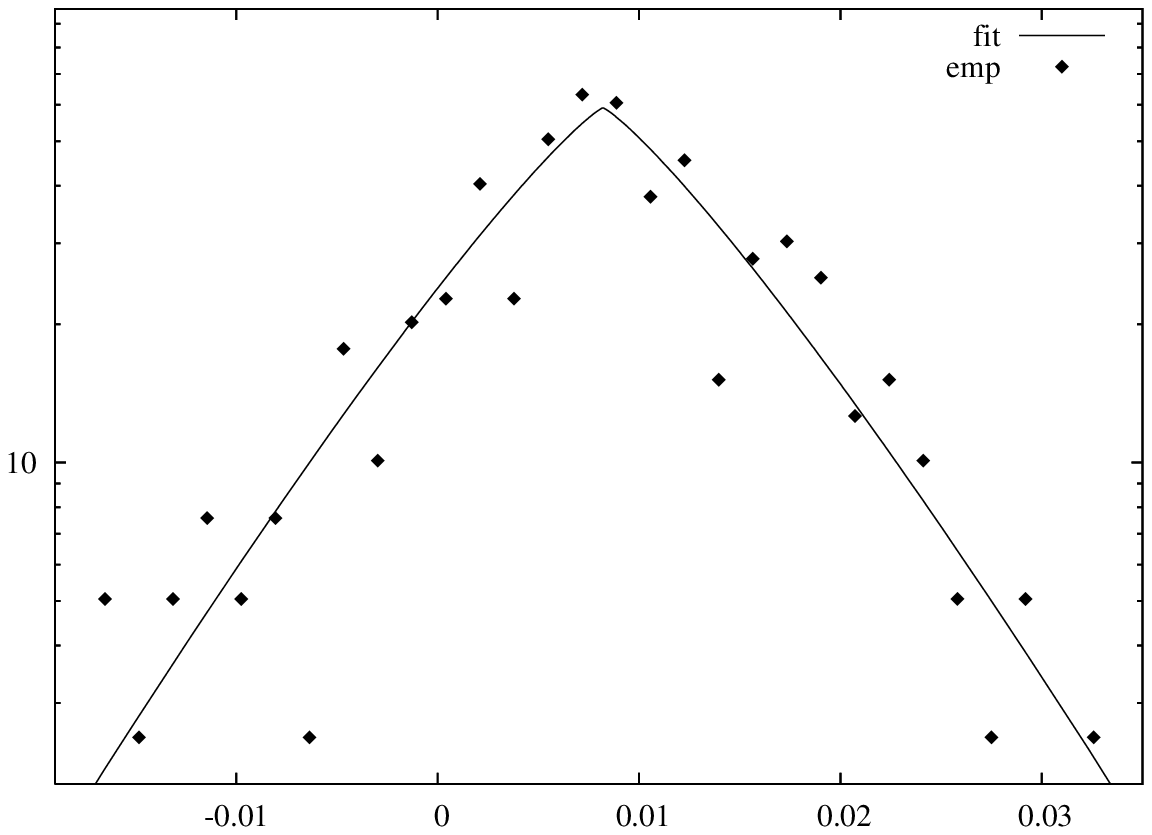}}
\caption{Binned Empirical Densities of U.S. GDP Growth Rates (emp)
vs. Subbotin Fit (fit).} \label{Fig:us_dy}
\end{minipage}\hfill
\begin{minipage}[h]{7.5cm}
\centering {\includegraphics[width=6cm]{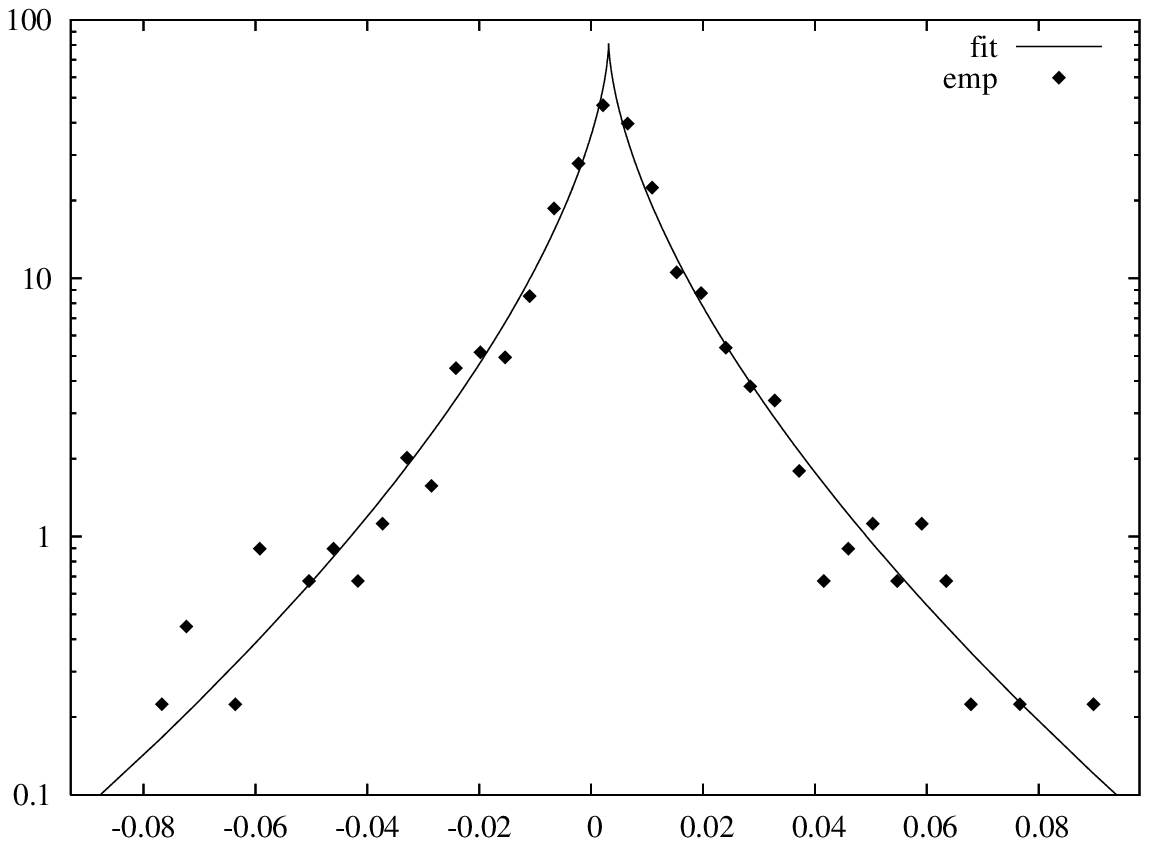}}
\caption{Binned Empirical Densities of U.S. IP Growth Rates vs.
Subbotin Fit. Time Period: $1921M1 - 2005M10$}\label{Fig:us_dip}
\end{minipage}
\end{figure}

Consider GDP first. In the first row of Table \ref{Tab:US}, we
show the maximum-likelihood estimates of Subbotin parameters and
their standard errors \footnote{The period of strong growth
experienced by the U.S. economy after World War II is probably
responsible for the positive location parameter $\widehat{m}$
(0.0082), which implies a positive sample average growth rate.}.
Estimates indicate that GDP growth rates seem to distribute
according to a Laplace: the shape parameter $\widehat{b}$ is equal
to 1.18, very close to the theoretical Laplace value of one.
Therefore, U.S. output growth rates display tails fatter than a
normal distribution. This can be also seen from Figure
\ref{Fig:us_dy}, where we plot the binned empirical density
vis-\`{a}-vis the fitted one \footnote{As one can visually
appreciate, the goodness-of-fit is quite good. Preliminary results
from extensive bootstrap goodness-of-fit testing exercises seems
to statistically support this visual evidence.}.

Next, we employ monthly industrial production (IP) as a proxy of
U.S. output \footnote{IP tracks very closely GDP in almost all
countries. More precisely, the GDP-IP cross-correlation profile
mimics from time $t-6$ to time $t+6$ the GDP auto-correlation
profile.}. Notice that, by focusing on IP growth, we can study a
longer time span and thus improve our estimates by employing a
larger number of observations. During the period $1921-2005$, the
IP growth rate distribution displays tails much fatter than the
Laplace distribution (see Fig. \ref{Fig:us_dip} and the $2nd$ row
of Table \ref{Tab:US}), an outcome probably due to the turmoils of
the Great Depression.

\begin{figure}[tbp]
\centering {\includegraphics[width=6cm]{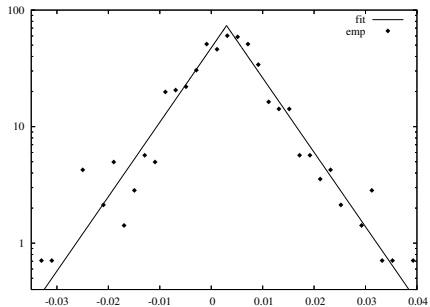}}
\caption{Binned Empirical Densities of U.S. IP Growth Rates vs.
Subbotin Fit. Time Period: $1947M1 - 2005M10$} \label{Fig:us_dips}
\end{figure}

Moreover, in order to better compare IP growth rate distribution
with the GDP one, we also carry out an investigation on the post
war period only ($1947-2005$). Notwithstanding this breakdown, our
results remain unaltered. In the post-war period the IP growth
rate distribution exhibits the typical ``tent-shape'' of the
Laplace density (cf. Fig. \ref{Fig:us_dips}). This outcome is
confirmed by a $\widehat{b}$ very close to one (see the third row
of Table \ref{Tab:US}). As pointed out by the lower standard
error, the estimate of $b$ is much more robust when we employ IP
series instead of the GDP one.

% Fig 4

\begin{figure}[tbp]
\centering {\includegraphics[width=12cm]{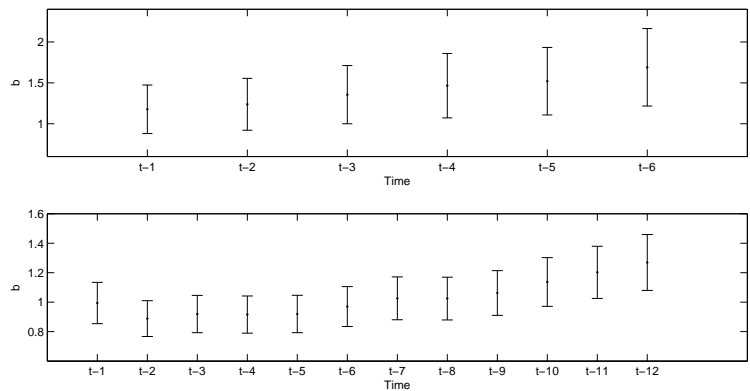}} \caption{U.S.
Output Growth Rates: Estimated Subbotin $b$ Parameter for
Different Time Lags. Error Bars (Cramer-Rao Bounds): $\pm
2\sigma(\widehat{b})$. Top Panel: GDP Data. Bottom Panel: IP Data.
} \label{Fig:bs}
\end{figure}

To perform a more precise check, one might also compute the
Cramer-Rao interval $[\widehat{b}-2
\sigma(\widehat{b}),\widehat{b}+2\sigma(\widehat{b})]$, where
$\sigma(\widehat{b})$ is the standard error of $\widehat{b}$
(Table \ref{Tab:US}, third column). A back-of-the-envelope
computation shows that, for all three growth rate series,
normality is always rejected. Moreover, one cannot reject the
Laplace hypothesis for both GDP and IP-1947 series, whereas tails
appear to be super-Laplace for IP-1921 \footnote{Cramer-Rao bounds
are also graphically reported in Figure \ref{Fig:bs} at lag
$t-1$.}.

Finally, in line with \cite{BnS06} we inspect the distribution of
output growth rates computed over longer lags. More precisely, we
consider growth rates now defined as:

\begin{equation}
g_{\tau}(t)=\frac{Y(t)-Y(t-\tau)}{Y(t-\tau)}\cong
y(t)-y(t-\tau)=dy_{\tau}(t), \label{eq:growth_rates_tau}
\end{equation}
where $\tau=1,2,...,6$ when GDP series is employed, and
$\tau=1,2,...,12$ when IP series is under study. In line with
\cite{BnS06}, we find that the shape parameter estimated on GDP
data becomes higher as $\tau$ increases (cf. the top panel of Fig.
\ref{Fig:bs}). When we consider IP series, the $\widehat{b}$ first
falls and then starts rising (see the bottom panel of Fig.
\ref{Fig:bs}). Therefore, as the ``growth lag'' increases, tails
become thinner (see \cite{Yako04} for similar evidence in the
contest of stock returns).

% Fig 6

\subsection{Cross-Country Analyses\label{Subsection:Cross}}

In the previous section we have provided evidence in favor of
fat-tailed (Laplace) U.S. output growth rate distributions. We now
perform a cross-country analysis in order to assess whether this
regularity pertains to the U.S. output only, or it might also be
observed in other developed countries. Our analysis focuses on the
following OECD countries: Canada, Japan, Austria, Belgium,
Denmark, France, Germany, Italy, Netherlands, Spain, Sweden, and
the U.K. .

\begin{table*}
\caption{Cross-Country Analysis of IP Time Series: Summary
Statistics. Asterisk (*): Significant at 5\% level.}
\begin{ruledtabular}
\begin{tabular}{lcccccc}
{\bf Series} & {\bf Mean} & {\bf Std. Dev.} & {\bf Skewness} & {\bf Kurtosis} & {\bf Jarque-Bera } & {\bf Lilliefors } \\
&  &  &  &   & {\bf  test} & {\bf  test} \\\hline
    Canada &     0.0021 &     0.0113 &    -0.2317 &     3.5631 &     5.9848 &     0.0391 \\
       USA &     0.0026 &     0.0073 &    -0.1505 &     4.6337 &   31.6281* &    0.0705* \\
     Japan &     0.0027 &     0.0404 &    -0.2250 &     4.6895 &   35.0981* &    0.0944* \\
   Austria &     0.0024 &     0.0253 &     0.1707 &     5.7806 &   90.8554* &    0.0565* \\
   Belgium &     0.0013 &     0.0401 &    -0.5689 &     5.9446 &  115.6987* &    0.0884* \\
   Denmark &     0.0025 &     0.0340 &     0.1214 &     7.2748 & 213.3210 &    0.0958* \\
    France &     0.0013 &     0.0130 &     0.1525 &     3.7251 &    6.9217* &    0.0740* \\
   Germany &     0.0015 &     0.0212 &     0.0098 &     9.2312 &  453.1891* &    0.0875* \\
     Italy &     0.0017 &     0.0321 &     0.0453 &     5.8380 &   93.3429* &    0.0692* \\
Netherlands &     0.0015 &     0.0285 &    -0.0350 &     6.5731 &  148.3145* &    0.0741* \\
     Spain &     0.0017 &     0.0401 &     0.2559 &     4.0067 &   14.5026* &     0.0469 \\
    Sweden &     0.0016 &     0.0302 &    -0.2955 &    37.0700 & 13627.2129* &    0.1153* \\
        UK &     0.0012 &     0.0140 &    -0.1631 &     8.4090 &  342.3813* &    0.0712* \\
\end{tabular}
\end{ruledtabular} \label{Tab:ss_cross}
\end{table*}

We start by analyzing the basic statistical properties of the
output growth rate time series (cf. Table \ref{Tab:ss_cross}). In
order to keep a sufficient time-series length, we restrict our
study to the industrial production series only. The standard
deviations of the IP series range from 0.0073 (U.S.) to 0.0404
(Japan). In half of the countries that we have analyzed, the
distributions of IP growth rates seem to be slightly right-skewed,
whereas in the other half they appear to be slightly left-skewed.
The analysis of the kurtosis reveals that in every country of the
sample the IP growth rate distribution is more leptokurtic than
the Normal distribution. Indeed, apart from Spain and Canada,
standard normality tests reject the hypothesis that IP growth
series are normally distributed.

Given this descriptive background, we turn to a country-by-country
estimation of the Subbotin distributions. Estimated coefficients
are reported in Table \ref{Tab:Cross}.

\begin{table}
\caption{Cross-Country Analysis of IP Time Series: Estimated
Subbotin Parameters.} \label{Tab:Cross}
\begin{ruledtabular}
\begin{tabular}{lcccccc}
{\bf } &                              \multicolumn{ 6}{c}{{\bf Estimated Parameters}} \\
    {\bf } & \multicolumn{ 2}{c}{{$\widehat{b}$}} & \multicolumn{ 2}{c}{{\bf $\widehat{a}$}} & \multicolumn{ 2}{c}{{\bf $\widehat{m}$}} \\
    {\bf Country} & {Par.} & {Std. Err.} & {Par.} & {Std. Err.} & {Par.} & {Std. Err.} \\
    \hline
    Canada &     1.6452 &     0.2047 &     0.0104 &     0.0007 &     0.0020 &     0.0010 \\
       USA &     1.2980 &     0.1516 &     0.0060 &     0.0004 &     0.0031 &     0.0004 \\
     Japan &     0.8491 &     0.0901 &     0.0259 &     0.0020 &     0.0021 &     0.0014 \\
   Austria &     1.2499 &     0.1446 &     0.0204 &     0.0014 &     0.0010 &     0.0014 \\
   Belgium &     1.0202 &     0.1125 &     0.0284 &     0.0021 &     0.0011 &     0.0017 \\
   Denmark &     0.8063 &     0.0847 &     0.0215 &     0.0017 &     0.0000 &     0.0012 \\
    France &     1.2623 &     0.1464 &     0.0106 &     0.0008 &     0.0010 &     0.0007 \\
   Germany &     0.9768 &     0.1067 &     0.0144 &     0.0011 &     0.0024 &     0.0008 \\
     Italy &     1.0778 &     0.1204 &     0.0237 &     0.0017 &     0.0010 &     0.0015 \\
Netherlands &     1.2133 &     0.1393 &     0.0223 &     0.0016 &     0.0019 &     0.0015 \\
     Spain &     1.4583 &     0.1755 &     0.0352 &     0.0024 &     0.0021 &     0.0029 \\
    Sweden &     0.8826 &     0.0944 &     0.0168 &     0.0013 &     0.0010 &     0.0009 \\
        U.K. &     1.0972 &     0.1230 &     0.0103 &     0.0008 &     0.0019 &     0.0006 \\
\end{tabular}
\end{ruledtabular}
\end{table}

The results of the cross-country analysis confirm that output
growth rates distribute according to a Laplace almost everywhere.
Excluding Canada, estimated ``shape'' coefficients are always
close to 1. If one considers the Cramer-Rao interval
$[\widehat{b}-2
\sigma(\widehat{b}),\widehat{b}+2\sigma(\widehat{b})]$, the only
country where output growth rate distribution does not appear to
be Laplace is Canada, whose $\widehat{b}$-interval lies above one
\footnote{Another exception is Denmark, where the upper bound of
its $\widehat{b}$-interval is slightly below one.}.

\subsection{Robustness Checks: Outliers, Heteroscedasticity, and Autocorrelation\label{Subsection:Cross}}

The foregoing discussion has pointed out that within-country
output growth rate distributions are markedly non-Gaussian. The
evidence in favor of Laplace (or super-Laplace) densities robustly
arises in the majority of OECD countries, it does not depend on
the way we measure output (GDP or IP), and it emerges also at
frequencies more amenable for the study of business cycles
dynamics (i.e. quarterly and monthly). Notice also that our
analysis does not show any clear evidence in favor of asymmetric
Laplace (or Subbotin) growth rate distributions. Hence, almost all
OECD countries seem to exhibit (with a probability higher than we
would expect) large, positive growth events with the same
likelihood of large, negative ones.

This ``fresh'' stylized fact on output dynamics must be however
scrutinized vis-\`{a}-vis a number of robustness checks. More
precisely, the above results can be biased by two classes of
problems. First, the very presence of fat-tails in the
distribution of country-level growth rates might simply be due to
the presence of outliers. Thus, one should remove such outliers
from the series and check whether fat tails are still there.
Second, our within-country analysis relies on pooling together
growth rate observations over time. Strictly speaking, the
observations contained in $G_{T_n}$ should come from i.i.d. random
variables. In other words, we should verify that fat tails do not
characterize growth rates only, but they are a robust feature of
growth residuals (also known as ``innovations''). To do so, one
might remove the possible presence of any structure in growth rate
time series due to autocorrelation and heteroscedasticity, and
then fit a Subbotin density to the residuals.

Our robustness analyses seem to strongly support the conclusion
that fat-tails still characterize our series also after having
controlled for outliers, autocorrelation and heteroscedasticity.
More precisely, in the first row of Table \ref{Tab:Rob} we have
reported the estimates of the Subbotin parameters in the case of
U.S. GDP, after having removed the most common types of outliers
\cite{DarDie2004}. The estimate for the shape parameter
($\widehat{b}$) still remains close to one, thus reinforcing
evidence in favor of Laplace fat-tails.

\begin{table*}
\caption{U.S. GDP Growth Rate Distribution: Estimated Subbotin
parameters after having removed outliers only (first row) and
after having removed both outliers and autocorrelation (second
row) from the original output growth rate series. Outlier removal
performed using TRAMO \cite{Tramo2001}. Autocorrelation removal
performed fitting an ARMA model to outlier-free residuals. Best
ARMA model: AR(1) w/o drift.}
\begin{ruledtabular}
\begin{tabular}{lcccccc}
{\bf }  & \multicolumn{ 6}{c}{{\bf Estimated Parameters}} \\
%\hline
    {\bf } & \multicolumn{ 2}{c}{{$\widehat{b}$}} & \multicolumn{ 2}{c}{{\bf $\widehat{a}$}} & \multicolumn{ 2}{c}{{\bf $\widehat{m}$}} \\
    {\bf After removing} & {Par.} & {Std. Err.} & {Par.} & {Std. Err.} & {Par.} & {Std. Err.} \\
\hline
  Outliers only &        1.2308 &     0.1568 &     0.0073 &     0.0006 &     0.0000 &     0.0006 \\
Outliers and autocorrelation &   1.2696 &     0.1628 &     0.0071 &     0.0006 &     0.0000 &     0.0006 \\

\end{tabular}
\end{ruledtabular}
\label{Tab:Rob}
\end{table*}

Moreover, in order to remove any structure from the growth rate
process, we have fitted a battery of ARMA specifications to the
growth rate time series obtained after cleanup of outliers and we
have selected the best model trough the standard Box and Jenkins's
procedure. In Table \ref{Tab:Rob}, second row, we report -- for
the case of U.S. GDP -- our Subbotin estimates for the
distribution of residuals of the best ARMA model, which turns out
to be an AR(1) without drift (thus implying the presence of some
autocorrelation in the original growth rate series). However, the
best fit for the distribution of the AR(1) residuals is a Subbotin
distribution very close to a Laplace ($\widehat{b}=1.2696$).
Similar results hold also for the IP growth rate series and are
reasonably robust across our sample of OECD countries.

Finally, we ran standard Ljung-Box and Engle's ARCH
heteroscedasticity tests on our growth series without detecting
any clear-cut evidence in favor of non-stationary variance over
time.

\begin{figure}[t]
\begin{minipage}[t]{7.5cm}
\centering \fbox{\includegraphics[width=6cm]{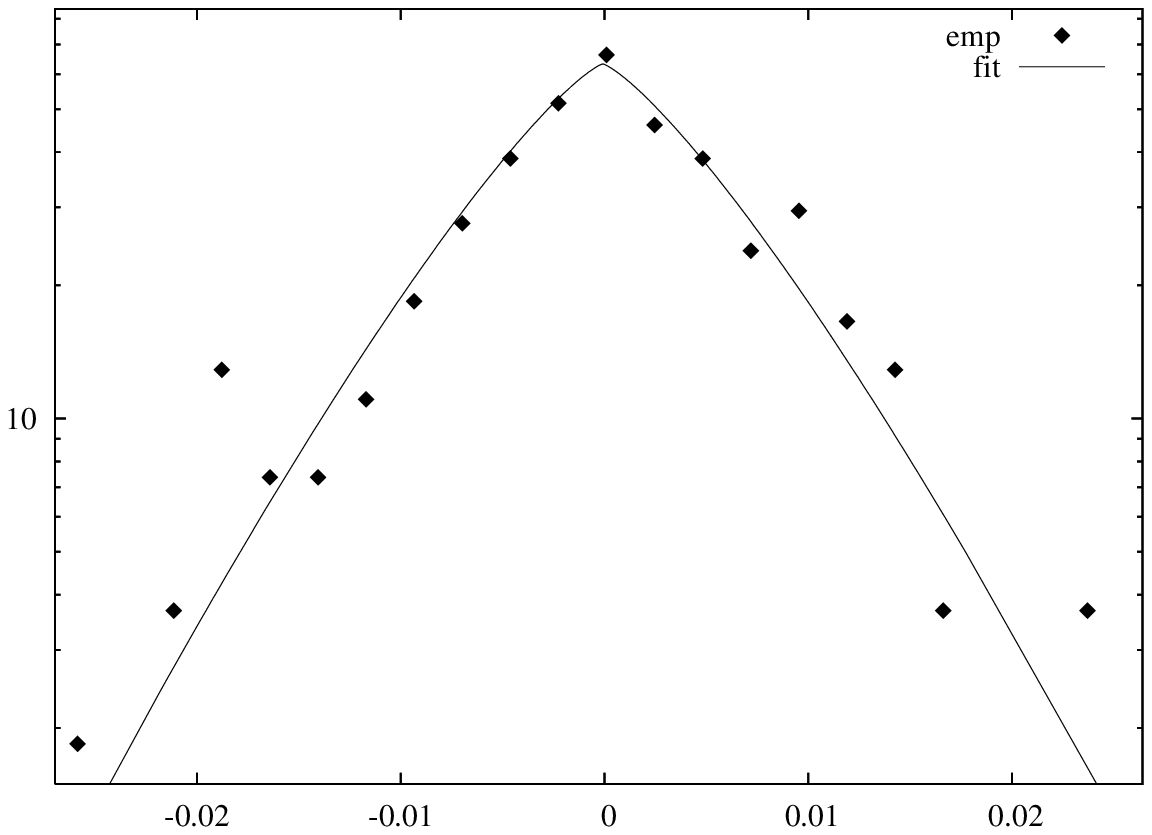}}
\end{minipage}\hfill
\begin{minipage}[t]{7.5cm}
\centering \fbox{\includegraphics[width=6cm]{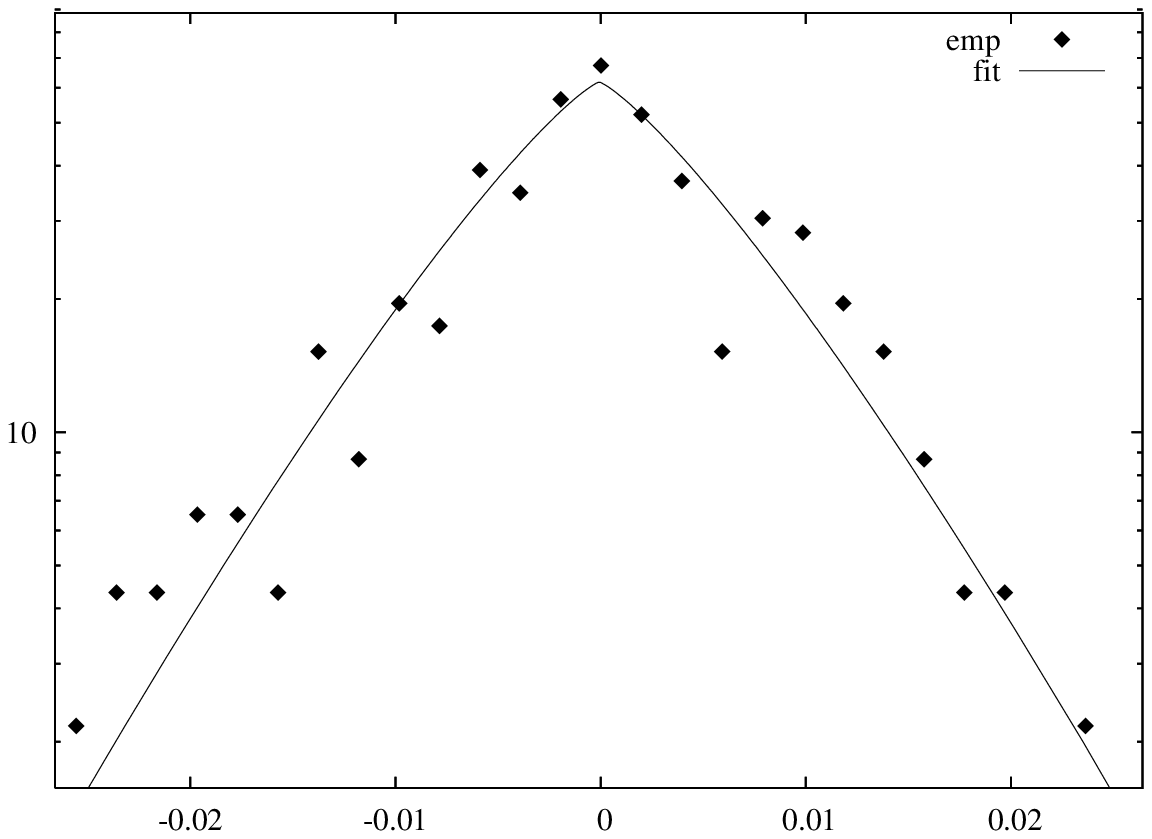}}
\end{minipage}
\caption{Controlling for outliers and autocorrelation in U.S.
Output Growth Rates. Binned Empirical Densities (emp) vs. Subbotin
Fit (fit). Left: Residuals after removing outliers only. Right:
Residuals after removing outliers and autocorrelation. Outlier
removal performed using TRAMO \cite{Tramo2001}. Autocorrelation
removal performed fitting an ARMA model to outlier-free residuals.
Best ARMA model: AR(1) w/o drift.} \label{Fig:tramo_arma}
\end{figure}

As Figure \ref{Fig:tramo_arma} shows for U.S. GDP, fat-tailed
Laplace densities seem therefore to robustly emerge even after one
washes away from the growth process both outliers and
autocorrelation (and moving-average) structure (i.e., when one
considers growth residuals as \textit{the} object of analysis).

\section{Concluding Remarks \label{Section:Conclusions}}

In this paper we have investigated the statistical properties of
GDP and IP growth rate time series distributions by employing
quarterly and monthly data from a sample of OECD countries.

We find that in the U.S., as well as in almost all other developed
countries of our sample, output growth rate time series distribute
according to a symmetric Laplace density. This implies that the
growth dynamics of aggregate output is \textsl{lumpy}, being
considerably driven by ``big events'', either positive or
negative. We have checked this result against a number of possible
sources of bias. We find that lumpiness appears to be a very
property of the data generation process governing aggregate output
growth, as it appears to be robust to the removal of both outliers
and auto-correlation.

At a very general and rather broad level, the robust emergence of
fat-tailed distributions for within-country time series of growth
rates and residuals can be interpreted as a fresh, new stylized
fact on output dynamics, to be added to the long list of its other
known statistical properties \footnote{Two rather undisputed
stylized facts of output dynamics -- at least for the U.S. -- are:
(i) GNP growth is positively autocorrelated over short horizons
and has a weak and possibly insignificant negative autocorrelation
over longer horizons; (ii) GNP appears to have an important
trend-reverting component
\cite{NnP82,Coc88,Coc94,BlaQuah89,Rud93,MnN00}.}.

>From a more empirical perspective, our results (together with the
already mentioned cross-section ones \cite{Sea96,
Cea98,Lea98,Aea97,BnS03a,BnS03b,CnD04,Fuetall05,Bea05,SnT06})
ought to be be interpreted together with recent findings against
log-normality for the cross-section distributions of firm and
country size \cite{Cab03,Dos05,Axt06,Quah96,Quah97}, and on
power-law scaling in cross-country per-capita GDP distributions
\cite{Gal03}. This joint empirical evidence seems to suggest that
in economics the room for normally-distributed shocks and growth
processes obeying the ``Law of large numbers'' and the ``Central
limit theorem'' is much more limited than economists were used to
believe. In other words, the general hint coming from this stream
of literature is in favor of an increasingly ``non-Gaussian''
economics and econometrics. A consequence of this suggestion is
that we should be very careful in using econometric testing
procedures that are heavily sensible to normality of residuals
\footnote{Such as Gibrat-like regressions for the dependence of
firm growth on size \cite{Sut97} and cross-section country growth
rates analyses \cite{Bar92}.}. On the contrary, testing procedures
that are robust to non-Gaussian errors and/or tests based on
Subbotin- or Laplace-distributed errors should be employed when
necessary.

Finally, country-level, non-Gaussian growth rates distributions
(both within-country and cross-section) might have an important
implication on the underlying generating processes. Suppose to
interpret the country-level growth rate in a certain time period
as the result of the aggregation of microeconomic (firm-level)
growth shocks across all firms and industries in the same time
period. The emergence of within-country non-Gaussian growth
distributions strongly militates against the idea that country
growth shocks are simply the result of aggregation of independent
microeconomic shocks over time. Therefore, some strong correlating
mechanism linking in a similar way at every level of aggregation
the units to be aggregated seems to be in place. This
interpretation is in line with the one proposed by
\cite{Aea97,Cea98,CnD04} who envisage the widespread presence of
fat tails as an indicator of the overall ``complexity'' of any
growth process, mainly due to the strong inner inter-relatedness
of the economic organizations under study.

\begin{acknowledgments}
Thanks to Giulio Bottazzi, Carolina Castaldi, Giovanni Dosi, Marco
Lippi, Sandro Sapio, Angelo Secchi and Victor M. Yakovenko, for
their stimulating and helpful comments. All usual disclaimers
apply.
\end{acknowledgments}

\bibliographystyle{apsrev}
\bibliography{subbobib}

\end{document}